# Final state boundary condition of the Schwarzschild black hole

Doyeol Ahn*

*Institute of Quantum Information Processing and Systems, University of Seoul, Seoul 130-743, Republic of Korea*

-**Abstract:** It is shown that the internal stationary state of the Schwarzschild black hole can be represented by a maximally entangled two-mode squeezed state of collapsing matter and infalling Hawking radiation. The final boundary condition at the singularity is then described by the random unitary transformation acting on the collapsing matter field. The outgoing Hawking radiation is obtained by the final state projection on the total wave function, which looks like a quantum teleportation process without the classical information transmitted. The black hole evaporation process as seen by the observer outside the black hole is now a unitary process but non-local physics is required to transmit the information outside the black hole. It is also shown that the final state projection by the evaporation process is strongly affected by the quantum state outside the event horizon, which clearly violates the locality principle.

*To whom correspondence should be addressed.
E-mail: dahn@uos.ac.kr ; davidahn@hitel.net

## I. Introduction

Hawking effect [1,2] on the information loss in black holes has been a serious challenge to modern physics because it requires a clear understanding of phenomena ranging from gravity to information theory. Hawking's semi-classical argument predicts that a process of black hole formation and evaporation is not unitary [3]. The vacuum state under the strong gravity of the black hole is composed of infalling and outgoing particle states inside and outside the event horizon. To an observer outside the black hole, the unitarity is lost because he would not be able to make any measurement in the interior of the event horizon and as a result he would be forced to make an average over the states in $H_{in}$ which corresponds to the Hilbert space inside of the horizon to obtain the density operator in $H_{out}$, the Hilbert space outside the black hole. On the other hand, there is some evidence in string theory that the formation and evaporation of black hole is a unitary process [4]. Nonetheless, Hawking effect, discovered nearly 30 years ago, is generally accepted very credible and considered as would be an essential ingredient of the yet unknown correct theory of quantum gravity.

Recently, Horowitz and Maldacena (HM) proposed a final-state boundary condition [4] in micro-canonical form to reconcile the unitarity of the black hole evaporation with Hawking's semi-classical reasoning. The essence of HM proposal is to impose a unique final boundary condition at the black hole singularity such that no information is absorbed by the singularity. The final boundary state is a maximally entangled state of the collapsing matter and infalling Hawking radiation. When a black hole evaporates, particles are created in entangled pairs with one falling into the black hole and the other radiated to infinity. The projection of final boundary state at the black hole singularity collapses the state into one associated with the collapsing matter and transfer the information to the outgoing Hawking radiation. The HM model is further refined, by including the unitary interactions between the collapsing matter and infalling Hawking radiation [5], and a random purification of the final boundary state [6]. The black hole evaporation process as seen by the observer outside the black hole is a unitary process, which looks like a quantum teleportation process [7] without the classical information transmitted. This indicates that non-local physics would be required to transmit the information outside the black hole and inside and outside the Hilbert spaces do not have independent existence. Then one could raise the following question: "In quantum theory which one is more fundamental, unitary or locality?" What locality requires is that there be no influence on an object due to any action taken in a region that is at space-like separation with respect to the object. If one forces the quantum gravity be unitary to observers outside the event horizon, then non-locality should be an essential feature of the theory [8].

In the HM model, the boundary state outside the event horizon is assumed to be the Unruh vacuum state [9, 10]. As a matter of fact, Hawking's original discovery can be regarded as imposing a boundary condition at the event horizon [11]. The author would like to denote it as Hawking boundary condition (HBC) in contrast with the final-state boundary condition (FBC) proposed by HM (Fig. 1). HBC dictates that the quantum states inside and outside the event horizon of the black hole are maximally entangled. The significance of HM proposal is that the black hole formation and evaporation process can be put into a unified picture by combining HBC together with FBC. Moreover, the process can be unitary as predicted by the string theory.

The author also showed that the black hole evaporation process will be affected by the boundary condition outside the event horizon [11]. Boundary state outside the event horizon affects the final state projection because the quantum states inside and outside the event horizon are entangled by HBC.

One of the critical assumptions in HM proposal is that the internal quantum state of the black hole can be represented by maximally entangled states of collapsing matter and infalling Hawking radiation. HM model is also based on the simplified micro-canonical form and it would be an interesting question how the final state boundary condition (FBC) will look like in a more general case such as Schwarzschild black hole.

In this paper, it is shown that the internal stationary state of the Schwarzschild black hole can be represented by a maximally entangled two-mode squeezed state of collapsing matter and infalling Hawking radiation. The outgoing Hawking radiation is obtained by the final state projection on the total wave function, which looks like a quantum teleportation process without the classical information transmitted. The black hole evaporation process as seen by the observer outside the black hole is a unitary process but non-locality is required to transmit the information outside the black hole. It is also shown that the final state projection by the evaporation process is affected by Hawking boundary condition (HBC) at the event horizon, which clearly violates the locality principle.

The purpose of this paper is to demonstrate that final state boundary condition

(FBC) of Horowitz and Maldacena necessarily implies a breakdown of locality in whatever quantum theory of gravity one might construct that incorporates this proposal. The author proceeds by (a) first reviewing the result that the original vacuum outside of a black hole evolves into a maximally entangled 2-mode squeezed state on $H_{in}$ and $H_{out}$ and then (b) shows that the interior state of the black hole is a maximally entangled 2-mode squeezed state on $H_{in}$ and $H_M$, where $H_M$ is the Hilbert space for the collapsing matter. The FBC is then applied to the latter state, and the outgoing radiation is obtained by projection of this onto the former state. However if one took an excited state outside of the black hole as the Hawking boundary condition then it too could be written as a maximally entangled 2-mode squeezed state on $H_{in}$ and $H_{out}$ - but the state of outgoing radiation that one would obtain via the FBC is orthogonal to what one obtains from the vacuum state (eq. (30)). Hence one obtains the result that the final outgoing particle state for black hole evaporation is dependent on the Hawking boundary condition. Since the interior and the exterior regions of the event horizon are causally disconnected, this is contrary to expectations based on locality. The author interprets this as indicating that non-locality is required to transmit information outside a black hole.

## II. Black hole final state in micro-canonical form

In this section, we review the original HM proposal briefly. We assume that the quantum state of the collapsing matter belongs to a Hilbert space $H_M$ with dimension $N$ and $|\psi\rangle_M$ be the initial quantum state of the collapsing matter. The Hilbert space of fluctuations on the background spacetime for black hole formation and evaporation is separated into $H_{in}$ and $H_{out}$ which contain quantum states localized inside and outside the event horizon, respectively. In HM proposal, HBC is assumed to be the Unruh vacuum state $|\Phi_0\rangle_{in \otimes out}$ belonging to $H_{in} \otimes H_{out}$ in micro-canonical form [4-6]:

$$|\Phi_0\rangle_{in \otimes out} = \frac{1}{\sqrt{N}} \sum_i |i\rangle_{in} \otimes |i\rangle_{out}, \qquad (1)$$

where $\{|i\rangle_{in}\}$ and $\{|i\rangle_{out}\}$ are orthonormal bases for $H_{in}$ and $H_{out}$, respectively. The final-state boundary condition (FBC) imposed at the singularity requires a maximally entangled quantum state in $H_M \otimes H_{in}$ which is called final boundary state and is given by

$$_{M \otimes in}\langle\Psi| = \frac{1}{\sqrt{N}} \sum_l {}_M\langle l| \otimes {}_{in}\langle l|(S \otimes I), \qquad (2)$$

where $S$ is a random unitary transformation. The initial matter state $|\psi\rangle_M$ evolves into a state in $H_M \otimes H_{in} \otimes H_{out}$ under HBC, which is given by $|\Psi_0\rangle_{M \otimes in \otimes out} = |\psi\rangle_M \otimes |\Phi_0\rangle_{in \otimes out}$. Then the transformation from the quantum state of collapsing matter to the state of outgoing Hawking radiation is given by the following final state projection [6]

$$|\phi_0\rangle_{out} = {}_{M \otimes in}\langle\Psi\|\Psi_0\rangle_{M \otimes in \otimes out} = \sum_i {}_M\langle i|S|\psi\rangle_M |i\rangle_{out}, \qquad (3)$$

where right side of Eq. (3) is properly normalized. Let's assume that the orthonormal bases $\{|i\rangle_{out}\}$ and $\{|l\rangle_M\}$ are related by the unitary transformation $T'$. The quantum state of the collapsing matter is transferred to the state of the outgoing Hawking radiation with fidelity

$$f_0 = \left|{}_{out}\langle\phi_0|T'|\psi\rangle_M\right|^2. \qquad (4)$$

I would like to note that we can also regard $T'$ as a tunnelling Hamiltonian [12] and the evaporation rate will be proportional to $\frac{2\pi}{\hbar} f_0$.

## III. Hawking radiation and gravitational collapse of a Schwarzschild black hole

### A. Hawking radiation

In this section, we first study the derivation of Hawking radiation by Unruh [9] and extend the results to the field of a collapsing matter inside the event horizon. The stationary Schwarzschild black hole is represented by the metric

$$ds^2 = -\left(1 - \frac{2M}{r}\right)dt^2 + \frac{dr^2}{1 - \frac{2M}{r}} + r^2\left(d\theta^2 + \sin^2\theta d\varphi^2\right), \tag{5}$$

where $M$ is the mass of the black hole. At $r = 2M$, the Schwarzschild spacetime has an event horizon. The general coordinate is $x^\mu = (t, r, \theta, \varphi)$ with the metric tensor given by

$$g_{tt} = -\left(1 - \frac{2M}{r}\right), \quad g_{rr} = \frac{1}{1 - \frac{2M}{r}}, \quad g_{\theta\theta} = r^2, \quad g_{\varphi\varphi} = r^2 \sin^2\theta. \tag{6}$$

The massless scalar field satisfies the wave equation

$$(-g)^{1/2} \frac{\partial}{\partial x^\mu}\left[g^{\mu\nu}(-g)^{1/2} \frac{\partial}{\partial x^\nu}\right]\phi = 0, \tag{7}$$

and the positive frequency normal mode solution is given by [9]

$$\phi_{\omega lm} = (2\pi|\omega|)^{-1/2} e^{-i\omega t} f_{\omega l}(r) Y_{lm}(\theta, \varphi), \tag{8}$$

where $f_{\omega l}(r)$ satisfies

$$\frac{\partial^2 f_{\omega l}}{\partial r^{*2}} + \omega^2 f_{\omega l} - \left(1 - \frac{2M}{r}\right)\left(\frac{l(l+1)}{r^2} + \frac{2M}{r^3}\right) f_{\omega l} = 0, \tag{9}$$

with $r^* = r + 2M\ln(r/2M - 1)$. If we denote the radiation part of the wave coming out of the past horizon of the black hole by $f_{\omega l}^-$, then it is give by

$$f_{\omega l}^-(r) \approx e^{i\omega r^*} + A_{\omega l}^- e^{-i\omega r^*}. \tag{10}$$

In Kruskal coordinate, the Schwarzschild metric becomes [13]

$$\begin{aligned}
ds^2 &= -2M \frac{e^{-r/2M}}{r} d\bar{u} d\bar{v} + r^2 d\theta^2 + r^2 \sin^2\theta d\varphi^2, \\
\bar{u} &= -4Me^{-u/4M}, \quad \bar{v} = 4Me^{v/4M}, \\
u &= t - r^*, \quad v = t + r^*, \\
r^* &= r + 2M\ln(r/2M - 1).
\end{aligned} \tag{11}$$

The Kruskal extension of the Schwarzschild spacetime is shown in Fig. 2. Since the Killing vector in Kruskal coordinate is given by $\partial/\partial\bar{u}$ on $H^-$ (Fig. 2), the solution in Kruskal coordinate is given by

$$\overline{\phi}_{\varpi lm} = (2\pi |\varpi|)^{-1/2} e^{-i\varpi \overline{u}} Y_{lm}(\theta,\varphi). \tag{12}$$

On the other hand, the original positive frequency normal mode on $H^-$ can be written by

$$\phi^-_{\omega lm} = (2\pi |\omega|)^{-1/2} (e^{-i\omega u} + A^-_{\omega l} e^{-i\omega v}) Y_{lm}(\theta,\varphi). \tag{13}$$

Using $e^{-i\omega u} = (|\overline{u}|/4M)^{i4M\omega}$ and $e^{-i\omega v} = (\overline{v}/4M)^{-i4M\omega}$ and the fact that $\overline{v} = 0$ on $H^-$ [13], we obtain

$$\phi^-_{\omega lm} = (2\pi |\omega|)^{-1/2} (|\overline{u}|/4M)^{i4M\omega} Y_{lm}(\theta,\varphi). \tag{14}$$

Since, $\overline{u} < 0$ in region $I$ and $\overline{u} > 0$ in region $II$ of the Fig. 2, the wave coming out of the past horizon of the black hole on $H^-$ can be written by

$$\phi^-_{\omega lm} = (e^{2\pi M\omega}{}_{out}\phi_{\omega lm} + e^{-2\pi M\omega}{}_{in}\phi_{\omega lm})/(2\sinh(4\pi M\omega))^{1/2}, \tag{15}$$

where ${}_{out}\phi_{\omega lm}$ vanishes inside the event horizon (region $II$) and ${}_{in}\phi_{\omega lm}$ vanishes in the exterior region of the black hole (region $I$). In Eq. (15), we have used the fact that $(-1)^{-i4M\omega} = e^{4\pi M\omega}$. Above definition of the positive frequency solution in terms of ${}_{out}\phi_{\omega lm}$ and ${}_{in}\phi_{\omega lm}$ leads to the Bogoliubov transformations [13,14] for the particle creation and annihilation operators in Schwarzschild and Kruskal spacetimes (Appendix A):

$$\begin{aligned} a_{K,\omega lm} &= \cosh r_\omega b_{out,\omega lm} - \sinh r_\omega b^\dagger_{in,\omega lm}, \\ a^\dagger_{K,\omega lm} &= \cosh r_\omega b^\dagger_{out,\omega lm} - \sinh r_\omega b_{in,\omega lm}, \\ \tanh r_\omega &= e^{-4\pi M\omega}, \quad \cosh r_\omega = (1-e^{-8\pi M\omega})^{-1/2}, \end{aligned} \tag{16}$$

where $a^\dagger_{K,\omega lm}$ and $a_{K,\omega lm}$ are the creation and annihilation operators, respectively, acting on the Kruskal vacuum in region $I$, $b^\dagger_{out,\omega lm}$ and $b_{out,\omega lm}$ are the creation and annihilation operators acting on the Schwarzschild vacuum of the exterior region of a black hole, and $b^\dagger_{in,\omega lm}$ and $b_{in,\omega lm}$ are the creation and annihilation operators acting on the Schwarzschild vacuum inside the event horizon. Then the ground state $|\Phi_o\rangle_{in\otimes out}$ which looks like the vacuum in the far past is a maximally entangled two-mode squeezed state on $H_{in} \otimes H_{out}$ [15-19] (Appendix B):

$$|\Phi_o\rangle_{in\otimes out} = \frac{1}{\cosh r_\omega} \sum_n e^{-4\pi M\omega n} |n\rangle_{in} \otimes |n\rangle_{out}, \tag{17}$$

where $\{|n\rangle_{in}\}$ and $\{|n\rangle_{out}\}$ are orthonormal bases (normal mode solutions) for $H_{in}$ and $H_{out}$, respectively. Equation (17) shows that the original vacuum state evolves to a two-mode squeezed state $|\Phi_o\rangle_{in\otimes out}$, which is also called the Unruh state [16-19], which resides on $H_{in} \otimes H_{out}$. This state contains a flux of outgoing particles in $H_{out}$. For the observer outside the black hole, the unitarity is lost because he would not be able to do any measurement in $H_{in}$ and as a result he would be forced to make an average over the states in $H_{in}$ to obtain the density operator in $H_{out}$. Unlike the case of a micro-canonical form, the Hilbert spaces are infinite dimensional.

**B. Gravitational collapse and black hole state**

Here, we show that the field inside the event horizon can be also decomposed into the collapsing matter field and the advanced wave incoming from infinity having similar form as the Hawking radiation thus paving a road to the final boundary condition. The Penrose diagram of a collapsing star [13] is shown in Fig. 3. The region *I* is a fragmentation of Fig. 2 including the region *II* (black hole). The collapsing shell metric in two-dimension is given by [9]

$$ds^2 = \begin{cases} -d\tau^2 + dr^2, & r < R(\tau) \\ -\left(1 - \frac{2M}{r}\right)dt^2 + \frac{dr^2}{1 - \frac{2M}{r}}, & r > R(\tau), \end{cases} \qquad (18)$$

with the shell radius $R(\tau)$ defined by

$$R(\tau) = \begin{cases} R_o, & \tau < 0 \\ R_o - v\tau, & \tau > 0. \end{cases} \qquad (19)$$

We define the advanced and retarded null coordinates as

$$\begin{aligned} V &= \tau + r - R_o, \quad U = \tau - r + R_o, \\ v^* &= t + r - R_o^*, \quad u^* = t - r^* + R_o^*, \end{aligned} \qquad (20)$$

with $R_o^* = R_o + 2M \ln(R_o/2M - 1)$. The null coordinates are chosen such that the shell begins to collapse at $U = V = u^* = v^* = 0$ [9]. After some mathematical manipulations, we obtain near the shell surface,

$$v^* \approx 4M \ln\left(1 - \frac{vV}{(1-v)(R_o - 2M)}\right), \qquad (21)$$

and

$$u^* \approx -4M \ln\left(1 - \frac{vU}{(1+v)(R_o - 2M)}\right). \qquad (22)$$

We now consider the massless scalar field inside the black hole, incoming from the infinity, which is given by

$$\phi^+_{\omega lm} = (2\pi |\omega|)^{-1/2} (e^{-i\omega v} - A^+_{\omega l} e^{-i\omega u}) Y_{lm}(\theta, \varphi), \qquad (23)$$

where $A^+_{\omega l}$ is chosen such that the field vanishes at $r = 0$. The normal mode on $H^+$ (Fig. 3) becomes

$$\phi^+_{\omega lm} = (2\pi |\omega|)^{-1/2} e^{i\omega R_o^*} \left| 1 - \frac{vV}{(1-v)(R_o - 2M)} \right|^{-i4M\omega} Y_{lm}(\theta, \varphi) \qquad (24)$$
$$= \left( e^{2\pi M\omega}{}_M\phi_{\omega lm} + e^{-2\pi M\omega}{}_{in}\phi_{\omega lm} \right)/(2\sinh(4\pi M\omega))^{-1/2},$$

where ${}_M\phi_{\omega lm}$ is a mode which vanishes for outside the shell, $V > (1-v)(R_o - 2M)/v$ and ${}_{in}\phi_{\omega lm}$ is the solution vanishing inside the shell, $V < (1-v)(R_o - 2M)/v$.

Above definition of the positive frequency solution in terms of ${}_M\phi_{\omega lm}$ and ${}_{in}\phi_{\omega lm}$ leads to the Bogoliubov transformation [13,14] for the particle creation and annihilation operators in Schwarzschild and Kruskal spacetimes as in the case of exterior region of the black hole:

$$c_{K,\omega lm} = \cosh r_\omega b_{M,\omega lm} - \sinh r_\omega b^\dagger_{in,\omega lm},$$
$$c^\dagger_{K,\omega lm} = \cosh r_\omega b^\dagger_{M,\omega lm} - \sinh r_\omega b_{in,\omega lm}, \qquad (25)$$
$$\tanh r_\omega = e^{-4\pi M\omega}, \quad \cosh r_\omega = (1 - e^{-8\pi M\omega})^{-1/2},$$

where $c^\dagger_{K,\omega lm}$ and $c_{K,\omega lm}$ are the creation and annihilation operators, respectively, acting on the Kruskal vacuum in region $II$, $b^\dagger_{M,\omega lm}$ and $b_{M,\omega lm}$ are the creation and annihilation operators for the collapsing matter acting on the Schwarzschild vacuum, and $b^\dagger_{in,\omega lm}$ and $b_{in,\omega lm}$ are the creation and annihilation operators acting on the Schwarzschild vacuum inside the event horizon. Then the stationary state $|\Phi_o\rangle_{M\otimes in}$ inside the black hole is a maximally entangled two-mode squeezed state on $H_M \otimes H_{in}$

$$|\Phi_o\rangle_{M\otimes in} = \frac{1}{\cosh r_\omega} \sum_n e^{-4\pi M\omega n} |n\rangle_M \otimes |n\rangle_{in}, \qquad (26)$$

where $\{|n\rangle_{in}\}$ and $\{|n\rangle_M\}$ are orthonormal bases (normal mode solutions) for $H_{in}$ and $H_M$, respectively.

## IV. Final state and Hawking boundary conditions

In the previous section, we have shown that the internal state for a fixed shell can be represented by the two-mode squeezed state of collapsing matter and infalling Hawking radiation. We now apply the final boundary condition at the singularity [4], which is given by the state $|\Psi\rangle_{M \otimes in}$:

$$_{M \otimes in}\langle\Psi| = {_{M \otimes in}}\langle\Phi_o|(S \otimes I)$$
$$= \frac{1}{\cosh r_\omega}\sum_n e^{-4\pi M\omega n}\left({_M}\langle n|S\right)\otimes {_{in}}\langle n|, \qquad (27)$$

where $S$ is a random unitary transformation [20]. The random unitary transformation S represents that the interior of the black hole is a turbulent place and it is difficult to distinguish the two subsystems presumed to compose the interior of the black hole [5]. The stationary state $|\Phi_o\rangle_{M \otimes in}$, a maximally entangled two-mode squeezed state of the infalling Hawking radiation and collapsing matter, derived by the semi-classical description corresponds to the state far from the singularity.

As in the case of the micro-canonical form, the initial matter state $|\psi\rangle_M \in H_M$, a pure state that will form the black hole, evolves into a state in $H_M \otimes H_{in} \otimes H_{out}$ under HBC, which is given by

$$|\Psi_0\rangle_{M \otimes in \otimes out} = |\psi\rangle_M \otimes |\Phi_0\rangle_{in \otimes out}$$
$$= |\psi\rangle_M \otimes \left(\frac{1}{\cosh r_\omega}\sum_n e^{-4\pi M\omega n}|n\rangle_{in} \otimes |n\rangle_{out}\right). \qquad (29)$$

The transformation from the quantum state of collapsing matter into the state of outgoing Hawking radiation when the black hole evaporates is given by the following final state projection [6]

$$|\phi_0\rangle_{out} = {_{M \otimes in}}\langle\Psi\|\Psi_0\rangle_{M \otimes in \otimes out}$$
$$= \frac{1}{\cosh^2 r_\omega}\sum_{n,m} e^{-4\pi M\omega(n+m)} {_M}\langle m|S|\psi\rangle_{M}\, {_{in}}\langle m\|n\rangle_{in} \otimes |n\rangle_{out} \qquad (30)$$
$$= \frac{1}{\cosh^2 r_\omega}\sum_n e^{-8\pi M\omega n} {_M}\langle n|S|\psi\rangle_M |n\rangle_{out}.$$

Eq. (30) shows that the black hole evaporation process is a unitary process from a pure state $|\psi\rangle_M$ in $H_M$ to another pure state $|\phi_0\rangle_{out}$ in $H_{out}$, which looks like a quantum teleportation process without the classical information transmitted. [21] On the other hand, this indicates that non-local physics would be required to transmit the information outside the black hole and inside and outside the Hilbert spaces do not have independent existence. To continue our discussion of the non-locality, we consider the change of HBC on the final evaporation process. If the quantum gravity is a local theory, then the final outgoing particle state at the black hole evaporation should be independent of the Hawking boundary condition because the interior and the exterior regions of the event horizon are causally disconnected. In the following, we show that this is not the case. We assume the Kruskal excited state as HBC, which is given by (Appendix B):

$$
\begin{aligned}
|\Phi_1\rangle_{in \otimes out} &= a^\dagger_{K,\omega lm} |\Phi_o\rangle_{in \otimes out} \\
&= \left( \cosh r_\omega b^\dagger_{out,\omega lm} - \sinh r_\omega b_{in,\omega lm} \right) \frac{1}{\cosh r_\omega} \sum_n e^{-4\pi M \omega n} |n\rangle_{in} \otimes |n\rangle_{out} \\
&= \frac{1}{\cosh^2 r_\omega} \sum_n e^{-4\pi M \omega n} \sqrt{n+1} |n\rangle_{in} \otimes |n+1\rangle_{out}.
\end{aligned}
\quad (31)
$$

Then the initial matter state $|\psi\rangle_M$ evolves into a state in $H_M \otimes H_{in} \otimes H_{out}$ under the new HBC, which is given by

$$
\begin{aligned}
|\Psi_1\rangle_{M \otimes in \otimes out} &= |\psi\rangle_M \otimes |\Phi_1\rangle_{in \otimes out} \\
&= |\psi\rangle_M \otimes \left( \frac{1}{\cosh^2 r_\omega} \sum_n e^{-4\pi M \omega n} \sqrt{n+1} |n\rangle_{in} \otimes |n+1\rangle_{out} \right),
\end{aligned}
\quad (32)
$$

and the state of outgoing Hawking radiation when the black hole evaporates given by the final state projection is

$$
\begin{aligned}
|\phi_1\rangle_{out} &= {}_{M \otimes in}\langle \Psi \| \Psi_1\rangle_{M \otimes in \otimes out} \\
&= \frac{1}{\cosh^3 r_\omega} \sum_{n,m} e^{-4\pi M \omega (n+m)} {}_M\langle m|S|\psi\rangle_{M \; in}\langle m \| n\rangle_{in} \otimes \sqrt{n+1} |n+1\rangle_{out} \\
&= \frac{1}{\cosh^3 r_\omega} \sum_n e^{-8\pi M \omega n} {}_M\langle n|S|\psi\rangle_M \sqrt{n+1} |n+1\rangle_{out}.
\end{aligned}
\quad (33)
$$

We would like to calculate the inner product between $|\phi_o\rangle_{out}$ and $|\phi_1\rangle_{out}$ to see how they are related to the final state projection:

$$_{out}\langle\phi_1\|\phi_o\rangle_{out} = \frac{1}{\cosh^5 r_\omega} \sum_{n,m} \sqrt{m+1} e^{-8\pi M\omega(n+m)} {}_M\langle\psi|S^\dagger|m\rangle_M {}_M\langle n|S|\psi\rangle_M {}_{out}\langle m+1\|n\rangle_{out}$$

$$= \frac{1}{\cosh^5 r_\omega} \sum_n \sqrt{n} e^{-8\pi M\omega(2n-1)} {}_M\langle\psi|S^\dagger|n-1\rangle_M {}_M\langle n|S|\psi\rangle_M \quad (34)$$

$$= {}_M\langle\psi|\left(\frac{1}{\cosh^5 r_\omega} \sum_n \sqrt{n} e^{-8\pi M\omega(2n-1)} S^\dagger|n-1\rangle_M {}_M\langle n|S\right)|\psi\rangle_M$$

$$= {}_M\langle\psi|\hat{Z}|\psi\rangle_M,$$

where an operator $\hat{Z}$ is defined by

$$\hat{Z} = \frac{1}{\cosh^5 r_\omega} \sum_n \sqrt{n} e^{-8\pi M\omega(2n-1)} S^\dagger|n-1\rangle_M {}_M\langle n|S. \quad (35)$$

From an elementary quantum theory, the term ${}_M\langle\psi|\hat{Z}|\psi\rangle_M$ in Eq. (34) is nothing but an expectation value of $\hat{Z}$ averaged over the initial matter state, which can also be obtained by the trace operation, which is given by

$$_M\langle\psi|\hat{Z}|\psi\rangle_M \cong tr(\hat{Z})$$

$$= tr\left(\frac{1}{\cosh^5 r_\omega} \sum_n \sqrt{n} e^{-8\pi M\omega(2n-1)} S^\dagger|n-1\rangle_M {}_M\langle n|S\right)$$

$$= tr\left(\frac{1}{\cosh^5 r_\omega} \sum_n \sqrt{n} e^{-8\pi M\omega(2n-1)} |n-1\rangle_M {}_M\langle n|SS^\dagger\right) \quad (36)$$

$$= tr\left(\frac{1}{\cosh^5 r_\omega} \sum_n \sqrt{n} e^{-8\pi M\omega(2n-1)} |n-1\rangle_M {}_M\langle n|\right)$$

$$= 0,$$

where we have used the fact that $tr(AB) = tr(BA)$ and $S$ is unitary. Eq. (36) shows that $|\phi_o\rangle_{out}$ and $|\phi_1\rangle_{out}$ are orthogonal states and the final state projection by the evaporation process is definitely affected by the Hawking boundary condition, which clearly violates the locality principle.

## V Summary

In summary, we have shown that the internal stationary state of the Schwarzschild black hole can be represented by a maximally entangled two-mode squeezed state of collapsing matter and infalling Hawking radiation. The final boundary condition at the singularity is then described by the random unitary transformation acting on the collapsing matter field. The outgoing Hawking radiation is obtained by the final state projection on the total wave function, which looks like a quantum teleportation process without the classical information transmitted. The black hole evaporation process as seen by the observer outside the black hole is now a unitary process but non-local physics is required to transmit the information outside the black hole. It is also shown that the final state projection by the evaporation process is definitely affected by the quantum state outside the event horizon, which clearly violates the locality principle.

**Acknowledgements** This work was supported by the Korea Science and Engineering Foundation, the Korean Ministry of Science and Technology through the Creative Research Initiatives Program under the contract No. R-16-1998-009-01001-0(2006). The author is also partially supported by the Brain Korea 21 from the Ministry of Education. D. A. thanks J. H. Bae for his help with figures.

**Appendix A: Derivation of equation (16)**

Let the quantum fields corresponding to the wave functions in Kruskal and Schwarzschild spacetimes be denoted as

$$_K\hat{\phi}_{\omega lm} = a_{K,\omega lm}\,{_K u_{\omega lm}} + a^\dagger_{K,\omega lm}\,{_K u^*_{\omega lm}}, \tag{A1}$$

and

$$\hat{\phi}^-_{\omega lm} = (e^{2\pi M\omega}\,{_{out}\hat{\phi}_{\omega lm}} + e^{-2\pi M\omega}\,{_{in}\hat{\phi}_{\omega lm}})/(2\sinh(4\pi M\omega))^{1/2}, \tag{A2}$$

with

$$\begin{aligned}_{out}\hat{\phi}_{\omega lm} &= b_{out,\omega lm}\,{_{out} u_{\omega lm}} + b^\dagger_{out,\omega lm}\,{_{out} u^*_{\omega lm}}, \\ _{in}\hat{\phi}_{\omega lm} &= b_{in,\omega lm}\,{_{in} u_{\omega lm}} + b^\dagger_{in,\omega lm}\,{_{in} u^*_{\omega lm}}.\end{aligned} \tag{A3}$$

Here $_\alpha u_\beta$ is a modal function. Quantum fields given by eqs. (A1) and (A2) should be equivalent on $H^-$ and as a result, we obtain

$$\begin{aligned}a_{K,\omega lm} &= \frac{e^{2\pi M\omega}}{(2\sinh(4\pi M\omega))^{1/2}} b_{out,\omega lm}({_K u_{\omega lm}},{_{out} u_{\omega lm}}) \\ &+ \frac{e^{2\pi M\omega}}{(2\sinh(4\pi M\omega))^{1/2}} b^\dagger_{out,\omega lm}({_K u_{\omega lm}},{_{out} u^*_{\omega lm}}) \\ &+ \frac{e^{-2\pi M\omega}}{(2\sinh(4\pi M\omega))^{1/2}} b_{in,\omega lm}({_K u_{\omega lm}},{_{in} u_{\omega lm}}) \\ &+ \frac{e^{-2\pi M\omega}}{(2\sinh(4\pi M\omega))^{1/2}} b^\dagger_{in,\omega lm}({_K u_{\omega lm}},{_{in} u^*_{\omega lm}}),\end{aligned} \tag{A4}$$

where $(\phi,\psi)$ is the Klein-Gordon inner product [10,13]. Calculations of Klein-Gordon inner products are straightforward and after some mathematical manipulations, we obtain [9]

$$\begin{aligned}a_{K,\omega lm} &= \frac{e^{2\pi M\omega}}{(2\sinh(4\pi M\omega))^{1/2}} b_{out,\omega lm} - \frac{e^{-2\pi M\omega}}{(2\sinh(4\pi M\omega))^{1/2}} b^\dagger_{in,\omega lm} \\ &= \cosh r_\omega\, b_{out,\omega lm} - \sinh r_\omega\, b^\dagger_{in,\omega lm},\end{aligned} \tag{A5}$$

with $\cosh r_\omega = (1 - e^{-8\pi M\omega})^{-1/2}$.

**Appendix B: Derivation of equations (17) and (31)**

Before we proceed, let's simplify the notation as follows:

$$a_{K,\omega lm} = a,\ b_{out,\omega lm} = b,\ b^\dagger_{in,\omega lm} = \bar{b}^\dagger,\ r_\omega = r. \tag{B1}$$

Then, eq. (A5) becomes

$$a = \cosh r\, b - \sinh r\, \bar{b}^\dagger, \tag{B2}$$

From the condition $a|\Phi_o\rangle_{in\otimes out} = 0$, we have

$$(\cosh r\, b - \sinh r\, \bar{b}^\dagger)|\Phi_o\rangle_{in\otimes out} = 0, \tag{B3}$$

or

$$(b - s\bar{b}^\dagger)|\Phi_o\rangle_{in\otimes out} = 0,\ s = \tanh r. \tag{B4}$$

Now, we assume that the Kruskal vacuum $|\Phi_o\rangle_{in\otimes out}$ is related to the Schwarzschild vacuum $|0\rangle_S$ by

$$|\Phi_o\rangle_{in\otimes out} = F(b, \bar{b}^\dagger)|0\rangle_S, \tag{B5}$$

where $F$ is some function to be determined later.

From $[b, b^\dagger] = 1$, we obtain $[b, (b^\dagger)^m] = \frac{\partial}{\partial b^\dagger}(b^\dagger)^m$ and $[b, F] = \frac{\partial}{\partial b^\dagger} F$. Then using equations (B4) and (B5), we get the following differential equation for $F$:

$$\frac{\partial F}{\partial b^\dagger} - s\bar{b}^\dagger F = 0, \tag{B6}$$

and the solution is given by

$$F \propto \exp(s b^\dagger \bar{b}^\dagger). \tag{B7}$$

By substituting (B7) into (B5) and by properly normalizing the state vector, we get

$$|\Phi_o\rangle_{in\otimes out} = (1-s^2)^{1/2} \sum_n s^n |n\rangle_{in} \otimes |n\rangle_{out}, \tag{B8}$$

which agrees with equation (17).

In the following, we derive eq. (31). From $a^\dagger = \cosh r\, b^\dagger - \sinh r\, \bar{b}$, we have

$$|\Phi_1\rangle_{in\otimes out} = a^\dagger |\Phi_o\rangle_{in\otimes out}$$

$$= \left(\cosh r\, b^\dagger - \sinh r\, \bar{b}\right)(1-s^2)^{1/2}\sum_n s^n |n\rangle_{in} \otimes |n\rangle_{out}$$

$$= (1-s^2)^{1/2}\cosh r \sum_{n=0} s^n \sqrt{n+1}|n\rangle_{in}\otimes|n+1\rangle_{out} - (1-s^2)^{1/2}\sinh r \sum_{n=1} s^n \sqrt{n}|n-1\rangle_{in}\otimes|n\rangle_{out}$$

$$= (1-s^2)^{1/2}\cosh r \sum_{n=0} s^n \sqrt{n+1}|n\rangle_{in}\otimes|n+1\rangle_{out} - (1-s^2)^{1/2}\cosh r \sum_{n=0} s^{n+2}\sqrt{n+1}|n\rangle_{in}\otimes|n+1\rangle_{out}$$

$$= (1-s^2)^{1/2}(1-s^2)\cosh r \sum_{n=0} s^n \sqrt{n+1}|n\rangle_{in}\otimes|n+1\rangle_{out}$$

$$= \frac{1}{\cosh^2 r}\sum_{n=0} \tanh^n r \sqrt{n+1}|n\rangle_{in}\otimes|n+1\rangle_{out},$$

where we have used the fact $\cosh r = (1-s^2)^{-1/2}$.

**Figure legends**

**Fig. 1.** Penrose diagram of for the black hole formation and evaporation processes [10]. A semi-classical theory which predicts the unitary evolution of black hole formation and evaporation has two boundary conditions: (1) Hawking boundary conditions (HBC) at the event horizon with quantum states in $H_{in}$ and $H_{out}$ maximally entangled and (2) the final-state boundary condition (FBC) inside the black hole for the quantum states of collapsing matter in $H_M$ and the infalling Hawking radiation in $H_{in}$ maximally entangled. $J_+$ and $J_-$ are future and past null infinity respectively.

**Fig. 2.** The Kruskal exptention of the Schwarzschild spacetime [10, 13]. In region *I*, null asymptotes $H_+$ and $H_-$ act as futue and past event horizons, respectively. The boundary lines labelled $J^+$ and $J^-$ are fute and past null infinities, respectively, and $i^o$ is the spacelike infinity.

**Fig. 3.** Penrose diagram of a collapsing star [13]. The region *I* is a fragmentation of Fig. 2 including the region *II* (black hole). The null ray $\gamma$ passes through the center of the collapsing matter and emerges to form the event horizon $H^+$.

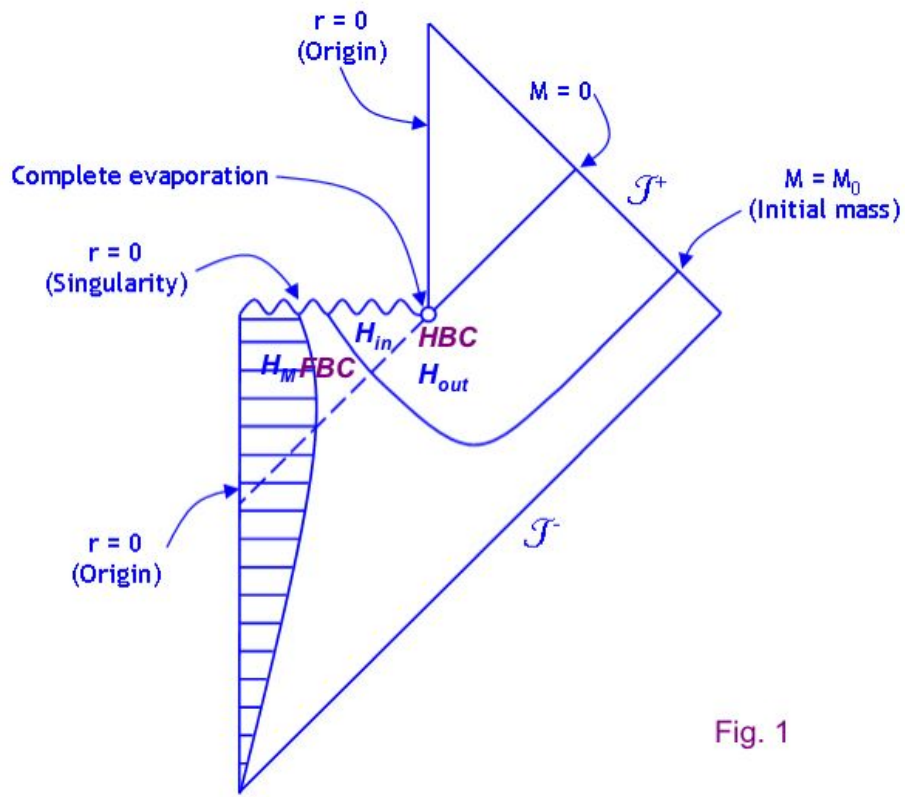

Fig. 1

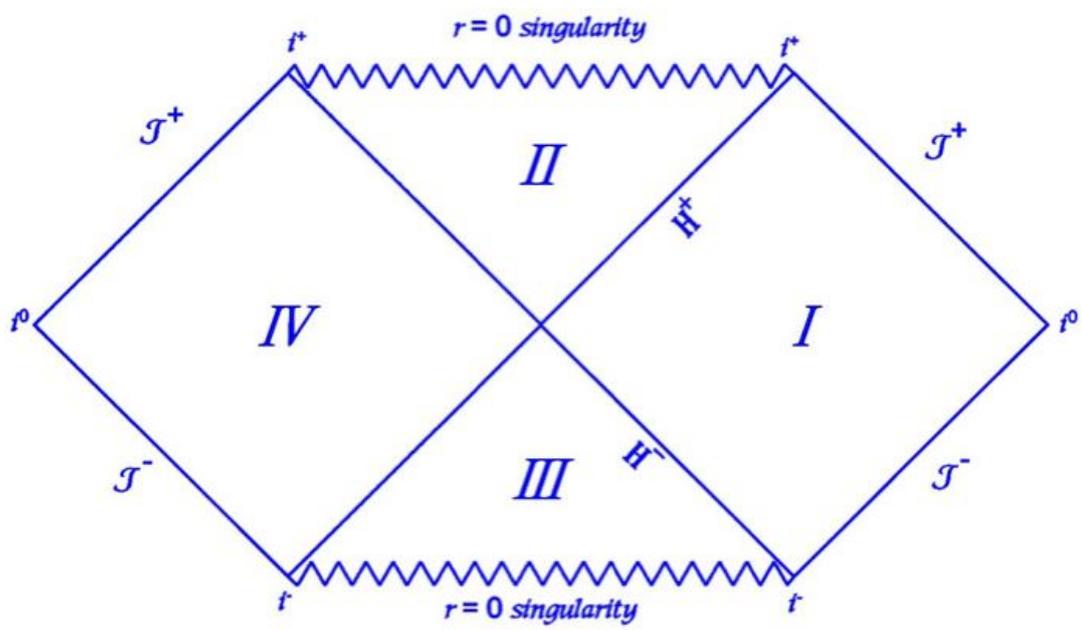

Fig. 2

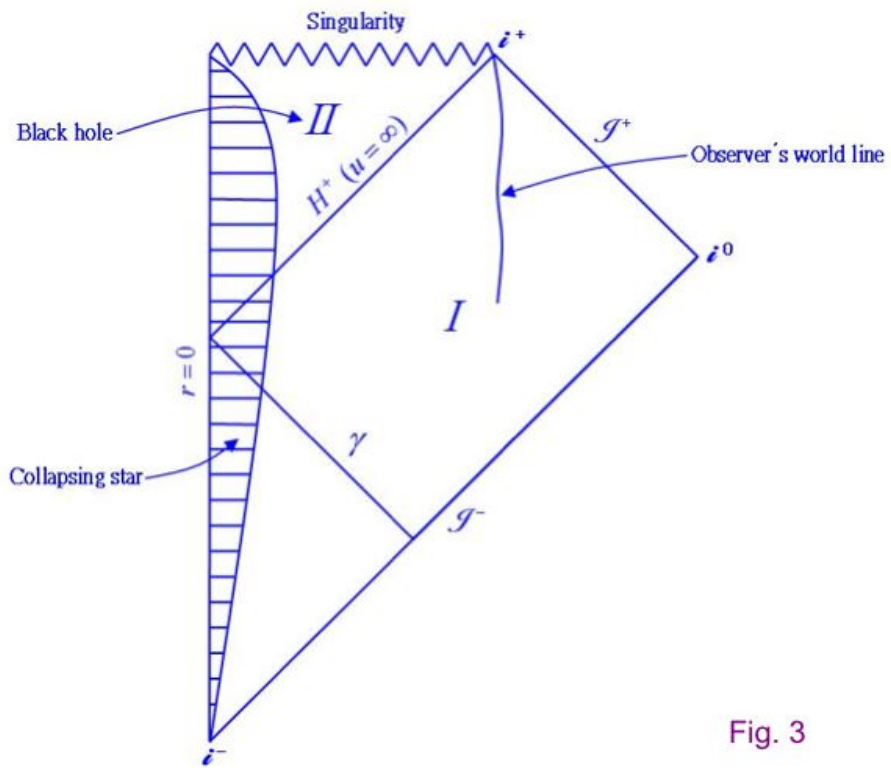

Fig. 3